\documentstyle[epsf]{mn}

\title{Line profile variations in $\gamma$ Doradus}
\author[L. A. Balona et al.]
{\Large L.A. Balona$^1$, T. B\"{o}hm$^2$, B.H. Foing$^3$, K.K. Ghosh$^4$,
 E. Janot-Pacheco$^5$,\\
{\Large K. Krisciunas$^6$, A-M Lagrange$^7$, W. A. Lawson$^8$, S. D.
James$^8$,  J. Baudrand$^9$,} \\
{\Large C. Catala$^{10}$, M. Dreux$^{10}$, P. Felenbok$^{10}$, J. B. Hearnshaw$^{11}$} \\
      $^1$South African Astronomical Observatory, P.O. Box 9, Observatory 7935,
Cape, South Africa\\
      $^2$ESO, Karl-Schwarzschildstr 2, D-85748, Garching, Germany\\
      $^3$ESA Space Science Department, ESTEC (SO), Postbus 299, 2200 AG
Noordwijk, The Netherlands\\
      $^4$Vainu Bappu Observatory, Indian Institute of Astrophysics,
Kavalur, Alangayam, T.N. 635701, India\\
      $^5$Departamento de Astronomia, Instituto Astronomico e 
Geofisico da USP, C.  Postal 9638, 01065-970 Sao Paulo, Brazil\\
      $^6$Joint Astronomy Centre, 660 N. A`ohoku Place, University Park,
Hilo, Hawaii 96720, USA\\
      $^7$Lab. Astrophys. Grenoble UJF, BP53X F, 38041 Grenoble Cedex,
France\\
      $^8$School of Physics, University College UNSW, Australian Defence
Force Academy, Canberra ACT 2600, Australia\\
      $^9$Dep. Astrophys. Extragal. et de Cosmologie, Obs. de
Paris, Section de Meudon, F-92195 Meudon Principal Cedex, France\\
      $^{10}$Obs. Midi-Pyrenees, 14 avenue Edouard Belin, 31400
Toulouse, France\\
      $^{11}$Mount John University Observatory, Department of Physics and
Astronomy, University of Canterbury, Christchurch, New Zealand}

\date{Accepted 1996 March 8.
      Received ........;
      in original form ........}

\begin{document}

\maketitle

\begin{abstract}
We present data from high-dispersion echelle spectra and simultaneous $uvby$
photometry for $\gamma$~Doradus.  These data were obtained from several
sites during 1994 November as part of the MUSICOS-94 campaign.  The star has
two closely-spaced periods of about 0.75 d and is the brightest member of
a new class of variable early F-type stars.  A previously suspected third
period, very close to the other two, is confirmed.  Previous observations
indicated that sudden changes could be expected in the spectrum, but none
were found during the campaign.  The radial velocities rule out the
possibility of a close companion.  The phasing between the radial velocity
and light curve of the strongest periodic component rules out the starspot
model.  The only viable mechanism for understanding the variability is
nonradial pulsation.  We used the method of moments to identify the modes
of pulsation of the three periodic components.  These appear to be
sectorial retrograde modes with spherical harmonic degrees, ($\ell, m$), as
follows: $f_1$ = (3,3), $f_2$ = (1,1) and $f_4$ = (1,1).  The angle of
inclination of the star is found to be $i \approx 70^\circ$.
\end{abstract}

\begin{keywords}
Stars: pulsation - Stars: variables - Stars: individual: $\gamma$~Dor.
\end{keywords}

\section{Introduction}

$\gamma$ Doradus (F0V) is the brightest member of what appears to be a new
class of pulsating variable stars.  In recent years, a number of late A- and
early F-type dwarfs have been found to be variable with periods of the order
of one day (see Krisciunas \& Handler, 1995 for a list).  This is too long for
membership of the $\delta$~Scuti class, where the periods are just a few hours.
The period is, in many cases, compatible with that expected for rotation,
which suggests that the variability may be due to the rotation of a spotted
star (Mantegazza, Poretti \& Zerbi 1994).  However, most stars appear to be
multiperiodic and require improbably large differential rotation.  The
starspot model for $\gamma$~Dor has been examined by Balona, Krisciunas
\& Cousins (1994b) who find that large overlapping spots would be required to
give the observed light and colour amplitudes.

The most likely explanation for the variability in these stars is nonradial
pulsation (NRP), since it is difficult to understand multiple periodicities
in terms of the starspot model or as orbital motion.  Multiple periods are
typical of NRP, but one needs to prove that the physical parameters
determined from an NRP model are realistic and that the other models are
unable to explain the observations.
In NRP, the light variability is due to the periodic variation in
temperature and, to a smaller extent, the change in the shape of the star
during pulsation.  Pulsation with a period of about one day implies that
the dominant restoring force is gravity ($g$-modes), rather than pressure
($p$-modes) as in the case of the $\delta$~Sct stars.  Furthermore, because
the stars show substantial light and radial velocity variations, the averaging
effect over the visible hemisphere of the star must be quite small.  This
implies that the spherical harmonic degree, $\ell$, of the pulsation must be
low, probably $\ell < 4$.  Aerts \& Krisciunas (1996) have analysed 9~Aur,
a star belonging to this group, using photometry and the cross-correlation
profile obtained with the CORAVEL instrument.  They deduce that 9~Aur is
pulsating in two modes, both having spherical harmonic indices ($\ell, m$) =
(3, $|1|$).

The mechanism which excites the pulsation is not known at present.  The
stars appear to be located at, or near, the red edge of the $\delta$~Sct
instability strip.  This indicates that the mechanism may be the same as for
$\delta$~Sct stars (the ionization of H and He), but perhaps modified by
convection.

It would certainly be interesting to examine the line profile and light
variations of these stars.  These data may enable us to distinguish between
NRP and rotational modulation of a spotted star.  NRP produces line profile
and light variations which, in principle, allow the determination of the
spherical harmonic degree, $\ell$ and the azimuthal number, $m$.  This can
be done by examining the variation of the moments of a line profile and
calculating a discriminant.  The value of the discriminant as a function
of angle of inclination is used to determine the most probable mode
($\ell, m$).  The method was developed by Balona (1986a, 1986b, 1987) and
extended by Aerts (1993).

The two periods of $\gamma$~Dor are stable and well determined from
extensive photometric observations: $P_1 = 0.75701$ d, $P_2 = 0.73339$ d
(Balona et al. 1994a).  More recently, evidence has been found for a third
period at $P_4 = 0.67797$ d (Balona, Krisciunas \& Cousins, 1994b).  (We
call this period $P_4$ and not $P_3$ because $P_3$ was used in Balona et al.
(1994a) to refer to a possible single period, later shown to be incorrect).  Since
the star is bright ($V = 4.3$) and well placed, it was chosen as one of the
targets for the MUSICOS 94 programme.  The aim was to obtain a large
number of high signal-to-noise line profiles which, in conjunction with
Str\"{o}mgren photometry, would decide which of the two models, starspots
or NRP, is the correct one and to determine the spot geometry or pulsation
modes.  Another factor in choosing to observe this star is the rapid change
in the spectrum which appears from time to time (Balona et al. 1994a).
$\gamma$~Dor is an {\it IRAS} source with a probable circumstellar dust cloud
(Aumann 1985), which may be responsible for this effect.

In this paper we present details of the observations and reduction procedure
and analyse the data in terms of the two models.

\section{Observations}

\subsection{Spectroscopy}

Spectroscopic data were obtained from four sites during 1994 November: ESO
(Chile), Mount Stromlo (Australia), Sutherland (South Africa) and Mount John
(New Zealand).  The projected rotational velocity of $\gamma$~Dor is high
($v \sin i = 50$ km s$^{-1}$), which means that it is impossible to find a
totally unblended spectral line.  The most suitable line is probably that of
Fe~II at $\lambda$5018.450 {\AA}  or possibly Mg~I at $\lambda$5183.619~{\AA}.
Instructions were given to ensure that the $\lambda$5018.450~Fe~II line was
observed whenever possible. The time-coverage at the various sites is shown
in Table 1.

\begin{table}
\caption{Details of the MUSICOS 94 campaign on $\gamma$~Dor.  The epoch of
the Julian day (JD) is 2449600.00; N is the number of spectra obtained.}
\begin{center}
\begin{tabular}{llrrr}\hline
  Site  & Observer & JD start &  JD end  &  N \\
\hline
ESO     & Lagrange & 59.58  & 59.88 & 80  \\
ESO     & Lagrange & 60.52  & 60.88 & 113 \\
ESO     & Lagrange & 62.52  & 62.87 & 88  \\
ESO     & Foing    & 63.58  & 63.88 & 90  \\
ESO     & Foing    & 64.60  & 64.88 & 113 \\
ESO     & Foing    & 65.61  & 65.87 & 79  \\
ESO     & Foing    & 66.61  & 66.90 & 99  \\
MSSSO    & Lawson   & 67.02  & 67.09 &  6  \\
MSSSO    & Lawson   & 67.92  & 68.05 &  5  \\
MSSSO    & Lawson   & 68.93  & 69.04 &  5  \\
MSSSO    & Lawson   & 70.91  & 71.13 & 15  \\
MSSSO    & Lawson   & 71.93  & 71.98 &  4  \\
MSSSO    & Lawson   & 72.91  & 73.13 & 13  \\
MSSSO    & Lawson   & 73.92  & 73.92 &  1  \\
MSSSO    & Lawson   & 74.92  & 75.19 & 14  \\
MtJohn  & Hearnshaw & 72.92 & 73.08 &  6 \\
MtJohn  & Hearnshaw & 74.90 & 75.00 &  5 \\
SAAO    & B\"{o}hm & 75.30  & 75.60 & 28  \\
SAAO    & B\"{o}hm & 76.31  & 76.59 & 25  \\
SAAO    & B\"{o}hm & 77.30  & 77.60 & 27  \\
SAAO    & B\"{o}hm & 78.28  & 78.42 & 11  \\
SAAO    & B\"{o}hm & 80.29  & 80.31 &  3  \\
SAAO    & B\"{o}hm & 81.28  & 81.29 &  2  \\
SAAO    & B\"{o}hm & 82.29  & 82.30 &  2  \\
SAAO    & B\"{o}hm & 83.29  & 83.45 &  5  \\
SAAO    & B\"{o}hm & 85.30  & 85.30 &  1  \\
\hline
\end{tabular}
\end{center}
\end{table}

The spectra from ESO were obtained using the Coud\'{e} Echelle Spectrometer
(CES) fed by the 1.4-m CAT. The detector was CCD No.9, a $1024 \times 640$
thinned back-illuminated RCA-type CCD.  The readout noise is about 25 e$^-$
(2 pixel on-line horizontal binning).  The grating efficiency was about 97
percent of maximum: the projected slit width was 48 $\mu$m.
Only one order was observed with spectral coverage 4998 -- 5035 {\AA}  and a
sampling of 0.033 {\AA}  pixel$^{-1}$.  The exposure times were in the range
3 -- 5 min, producing well-exposed spectra near the saturation limit.  The
resolving power was $R = 40 000$.  A Th-Ar lamp was used for wavelength
calibration.  An example of the spectrum, normalised to the continuum
(estimated by eye), is shown in Fig.~1.

\begin{figure}
\epsfxsize=8.3cm
\epsffile[58 79 448 270]{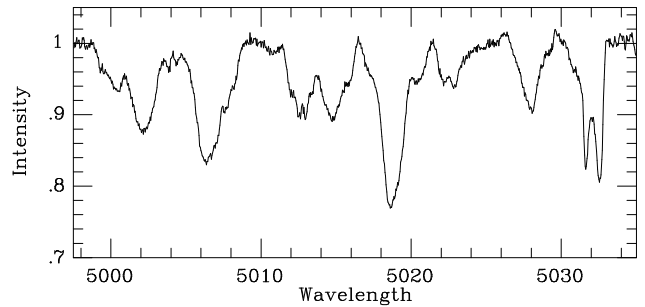}
\caption{An example of the normalised spectrum of $\gamma$~Dor obtained
at ESO.  The feature at 5032 {\AA}  is an artifact; the spectrum has not
been smoothed.}
\end{figure}

The Mt Stromlo and Siding Springs Observatories (MSSSO) spectra were obtained 
with the coud\'{e} echelle on the 1.9-m telescope. The echelle grating has
79 lines mm$^{-1}$ with the cross-disperser blazed at 5000 {\AA}.  The detector
was a $2048 \times 2048$ Tek CCD with gain of 2 e$^-$/ADU and readout noise
of about 2600 e$^-$ (no on-chip binning).  The slit width was 200 $\mu$m
(0.8 arcsec on the sky or 2 pixels on the detector).  The pixel size was about
25 $\mu$m and resolving power $R \approx 60 000$.  A Th-Ar lamp was used for wavelength calibration.  The full
spectral range of each order was covered. Each order covers about 87 {\AA}
(0.043 {\AA}  pixel$^{-1}$) and overlaps with neighbouring orders.  An example
of a spectrum obtained from 12 orders in the best-exposed range
(4900 -- 5500 {\AA}) is shown in Fig.~2.

\begin{figure}
\epsfxsize=8.3cm
\epsffile[56 55 421 591]{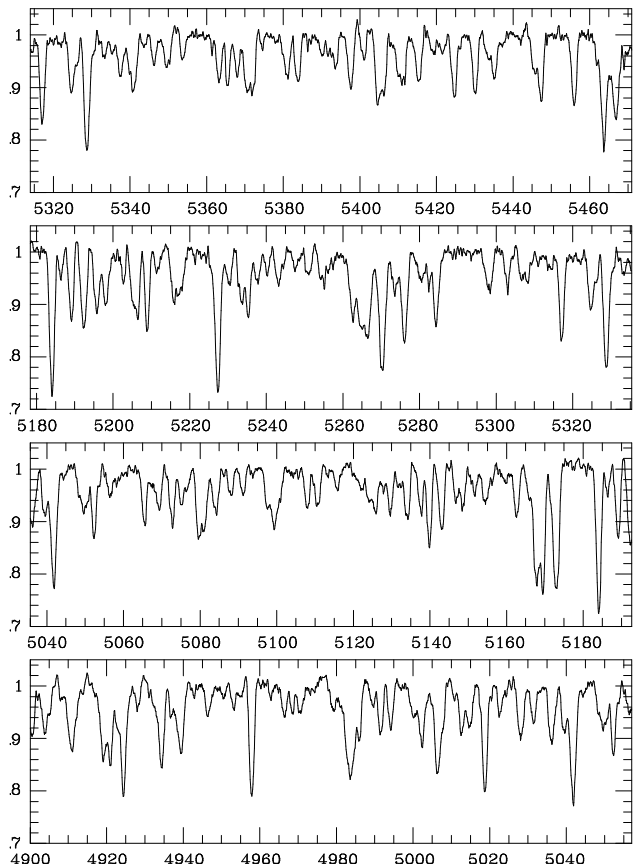}
\caption{An example of the normalised spectrum of $\gamma$~Dor obtained
from several echelle orders at MSSSO.  Spectra from SAAO look almost
identical.  A running mean box filter with a full width of seven pixels has
been applied.}
\end{figure}

The South African Astronomical Observatory (SAAO) spectra were obtained 
with the MUSICOS spectrograph attached to the 1.9-m telescope at Sutherland.
Details of the instrumentation can be found in Baudrand \& B\"{o}hm (1992).
The detector was a $1024 \times 1024$ Tek CCD with gain of 2 e$^-$/ADU and
readout noise of 46 e$^-$ (no on-line binning).  The pixel size was 24
$\mu$m; the spectral resolution $R = 31 000$.    The full spectral range
of each order was covered, neighbouring orders overlap in wavelength.  Each
order covered about 76 {\AA}  (0.074 {\AA}  pixel$^{-1}$).  The resulting spectra
are very similar to those obtained as MSSSO (Fig.~2).  A Th-Ar calibration
lamb was used.

Spectra from Mt John University Observatory were obtained 
using the 1.0-m McLellan reflector and a Thomson TH 7882 CDA CCD as
detector.  This chip has $384 \times 576$ pixels of dimension 23 $\mu$m.
The slit width was 100 $\mu$m, corresponding to 2.0 arcsec on the sky and
giving a wavelength resolution of about 0.2 {\AA}  and a resolving power $R =
25 000$.  Details of the spectrograph are given in Hearnshaw (1977) and of
the CCD detector system in Tobin (1992).  These spectra have lower  signal
to noise ratio than the others and a running mean box filter with full
width of 15 pixels was applied.  Neighbouring orders do not overlap, so it
is not possible to construct a continuous spectrum as was done for the Mt
Stromlo and SAAO data.  The Mt John data were consequently used only to
study the $\lambda$5018.450 {\AA}  line of Fe~II and to monitor possible rapid
changes in this and other spectral regions.

\subsection{Photometry}

Str\"{o}mgren $uvby$ photometry was obtained by L Balona during the last two
weeks of 1994 November (simultaneously with the MUSICOS run at Sutherland)
using the single-channel photometer attached to the 0.5-m reflector at
Sutherland.  During this period, K Krisciunas obtained Johnson $V$ photometry
from CTIO using the 0.6-m Lowell reflector.  The
weather was very good from both sites: 12 out of 14 nights were completely
or partly photometric at Sutherland, while all 10 nights were photometric at
CTIO.  A total of 132 $uvby$ and 245 $V$ observations were obtained.  The
coverage was 10 -- 18 observations per night at Sutherland and 20 -- 30
observations per night at CTIO.  In both cases a neutral density filter had
to be used to avoid saturating the photomultiplier.  The nearby F-type
stars, HR~1291 and HR~1365 were used as local comparison stars. 
Observations of these two stars flanked each observation of $\gamma$~Dor.

The spectroscopic and photometric data used in this paper can be obtained by
examining the World Wide Web home page of the SAAO ({\tt www.saao.ac.za}).

\section{Spectroscopic reductions}

Reductions of the Mt John data were made using the MIDAS
software package.  All other data were reduced using {\tt SPEC1} (single-order
spectra) and {\tt SPEC2} (multi-order echelle spectra).  These programs were
written in {\tt FORTRAN} by L Balona.  The programs require, for each night,
a list of files of three types: flatfield, arc and object frames.  A few
object frames are co-added and used to define the ridge line of the
spectrum as a function of $(X, Y)$ position on the CCD.  The flatfield
frames can also be used for this purpose.  A certain number of pixels on
either side of the ridge line are used to define the extent of the spectrum
perpendicular to the dispersion axis (the $X$-axis).  These were chosen so
as to include practically all the light from the star.  A one-dimensional
spectrum is formed by adding the pixels along the $Y$ direction inside the
strip for a given $X$ and subtracting the background for this value of $X$.
The background is obtained from two strips on either side of the ridge line
and sufficiently distant from it as to exclude any starlight.  The two strips
are added and a running mean applied to define the background as a function
of $X$.

The flatfield spectrum is obtained by co-adding all the flatfield frames
(excluding those with saturated pixels) and extracting a strip coincident
with the ridge line of the stellar spectrum and of the same width.  The pixels
along the $Y$ direction inside the strip are added and the background
subtracted to form a one-dimensional spectrum.  The resulting flatfield
spectrum is normalised to unit mean and applied to the object spectra.

The arc frames are co-added and a strip coincident with the ridge line of
the stellar spectrum, and of the same width, is extracted.  This is
converted to a one-dimensional spectrum and the background subtracted as
described above.  Finally, the flatfield correction is applied.  The arc
lines are fitted by Gaussians using non-linear least
squares.  This gives precise $X$ positions, which, together with the
laboratory wavelengths, are used to calibrate wavelength as a function of
$X$.  The arc lines are identified automatically by supplying the approximate
wavelengths at two points on the spectrum as well as a list of wavelengths
for the fitted lines.  This information is contained in the configuration
file.  A polynomial of second-order was found sufficient to fit the
calibration curve with a precision of 0.02 -- 0.05 {\AA}  rms.
The output of {\tt SPEC1} is a list of ASCII files, one for each object frame,
giving intensity versus wavelength.

{\tt SPEC2} works in the same way as {\tt SPEC1}, but on as many orders as desired.  As
input, the approximate $Y$ positions of the centre of each order is
required.  Output is a binary file containing the intensity versus
wavelength of each order.  Subsequent programs read these data and calculate
the required intensity factors to match the orders (using the wavelength
overlap between orders).  Spikes are removed in the process to produce a
continuous spectrum covering many orders.  Finally, the continuum is estimated
by eye and fitted to several points using a spline curve.  This is used to
normalise the spectrum.  For consistency, the continuum points were chosen
to be at the same wavelengths for all spectra.  The Mt Stromlo spectra
suffer from a misaligned flatfield, which produces a rather bumpy
continuum for each order when the flatfield is applied.  The normalisation
technique just described was successful in correcting this problem.  For
both the Mt Stromlo and SAAO data we applied a box running mean
filter with total width of seven pixels to remove excess noise (though the
unsmoothed spectra could be used).

\section{Period analysis}

The periodogram of the combined SAAO $y$ and CTIO $V$ photometry is shown in
Fig.~3.  The time span of 13 d is not sufficient to resolve the two
closely-spaced frequencies.  The single peak at $f = 1.342$ d$^{-1}$ is a
mean of $f_1 = 1.32098$ and $f_2 = 1.36354$ d$^{-1}$.  If $f_1$ is fitted to
the data, a periodogram of the residuals has the strongest peak at 1.387 
d$^{-1}$.  Conversely, if $f_2$ is fitted and removed from the data, the
strongest peak in the periodogram is at 1.303 d$^{-1}$.  Because the total
duration of the observations is too short to resolve $f_1$ and $f_2$, it is
not possible to obtain accurate values of the frequencies.  Nevertheless,
the fact that prewhitening by the known value of $f_1$ produces a peak close
to $f_2$ and vice-versa, shows that the two frequencies are still present.

\begin{table*}
\begin{minipage}{17.5cm}
\caption{Amplitudes ($A_j$) and phases ($\phi_j$) of a three-component
Fourier fit to the $V$ data between 1989 and 1994.  The frequencies are
$f_1 = 1.32098$, $f_2 = 1.36354$, $f_4 = 1.47447$ d$^{-1}$ and the epoch
is JD 2449000. Semi-amplitudes are in millimags; phases are in fractions
of a period.  The JD is with respect to 2440000; $N$ is the number of
observations.}
\begin{center}
\begin{tabular}{rrrrrrrr}\hline
 JD & $N$
 & \multicolumn{1}{c}{$A_1$} &  \multicolumn{1}{c}{$\phi_1$} 
 & \multicolumn{1}{c}{$A_2$} &  \multicolumn{1}{c}{$\phi_2$} 
 & \multicolumn{1}{c}{$A_4$} &  \multicolumn{1}{c}{$\phi_4$} \\
\hline
7838 - 7970 & 175  &  $ 16.8 \pm 1.0$ & $  0.03 \pm 0.01$
                   &  $ 12.9 \pm 1.0$ & $  0.26 \pm 0.02$
                   &  $  3.0 \pm 1.0$ & $  0.56 \pm 0.05$ \\
8185 - 8327 & 107  &  $ 12.8 \pm 1.2$ & $  0.04 \pm 0.02$
                   &  $ 15.6 \pm 1.2$ & $  0.27 \pm 0.01$
                   &  $  2.6 \pm 1.2$ & $  0.39 \pm 0.08$ \\
8993 - 9005 & 278  &  $ 16.3 \pm 1.2$ & $  0.09 \pm 0.01$
                   &  $ 19.5 \pm 1.5$ & $  0.31 \pm 0.01$
                   &  $  2.6 \pm 0.3$ & $  0.26 \pm 0.03$ \\
9326 - 9406 & 331  &  $  6.8 \pm 0.7$ & $  0.98 \pm 0.02$
                   &  $  9.5 \pm 0.7$ & $  0.19 \pm 0.01$
                   &  $  5.9 \pm 0.6$ & $  0.91 \pm 0.02$ \\
9672 - 9685 & 377  &  $ 14.9 \pm 0.7$ & $  0.92 \pm 0.01$
                   &  $ 13.4 \pm 0.7$ & $  0.21 \pm 0.08$
                   &  $  7.2 \pm 0.5$ & $  0.14 \pm 0.01$ \\
\hline
\end{tabular}                                                    
\end{center}
\end{minipage}
\end{table*}

Since $f_1$ and $f_2$ are still present, it is reasonable to assume that the
frequencies have remained unchanged over the last few years.  To test this,
we show in Table~2 the amplitudes and phases for individual seasons using
$f_1 = 1.32098$ and $f_2 = 1.36354$ d$^{-1}$ (the third frequency, $f_4$, is
discussed below).  It can be seen that the phases over the five years do not
differ significantly from the mean, in spite of the small formal errors
(these do not take into account the distribution of the observations).  The
maximum deviation in phase is 0.09 periods for $f_1$ and 0.06 periods for
$f_2$, while the standard deviation is about 0.03 periods in both cases. 
The low amplitude during the 1993 season is particularly striking.

\begin{figure}
\epsfysize=8.3cm
\epsffile[10 79 286 427]{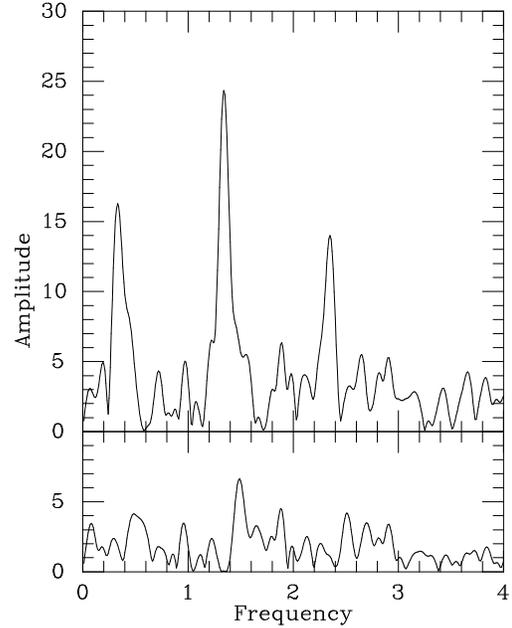}
\caption{Periodogram of the combined SAAO $y$ and CTIO $V$ data of
$\gamma$~Dor (top panel) and the periodogram after removal of the two principal components, $f_1$ and
$f_2$ (bottom panel).  Amplitude is in mmag, frequency in cycles d$^{-1}$.}
\end{figure}

Fig.~3 also shows the periodogram of the 1994 November $y$ and $V$ data after
removing the two periodic components.  It is clear that a third frequency at $f =
1.4888$ d$^{-1}$ is present.  This was strongly suspected by Balona, Krisciunas \&
Cousins (1994b) who obtained $f_4 = 1.475$ d$^{-1}$, $A_4 = 7.4$ mmag.  This
third frequency, $f_4$, is therefore confirmed.  Table~2 shows that the
amplitude of this component was below the detectable threshold (about 3
mmags) prior to the 1993.  A periodogram of the combined 1993 and 1994 seasons
gives a best estimate of $f_4 = 1.47447 (\pm 5)$ d$^{-1}$.  Table 3 lists the
amplitudes and phases of the three periodic components of the $V$ and
Str\"{o}mgren colour indices for the combined 1993 and 1994 seasons.  This
gives the best estimate of the phases for 1994.  However, the amplitudes of
$f_1$ and $f_2$ for 1994 should probably be taken from Table~2, since they
were particularly low in 1993.

\begin{table*}
\begin{minipage}{17.5cm}
\caption{Amplitudes ($A_j$) and phases ($\phi_j$) of a three-component
Fourier fit to the radial velocity, $V_r$ and $V$, $b-y$, $c_1$ and $u-b$
(1993 and 1994 seasons).  The frequencies are $f_1 = 1.32098$,
$f_2 = 1.36354$, $f_4 = 1.47447$ d$^{-1}$ and the epoch is JD 2449000.
Semi-amplitudes are in km s$^{-1}$ for $V_r$ and in millimags for the
photometry; phases are in fractions of a period.}
\begin{center}
\begin{tabular}{crrrrrr}\hline
 & \multicolumn{1}{c}{$A_1$} &  \multicolumn{1}{c}{$\phi_1$} 
 & \multicolumn{1}{c}{$A_2$} &  \multicolumn{1}{c}{$\phi_2$} 
 & \multicolumn{1}{c}{$A_4$} &  \multicolumn{1}{c}{$\phi_4$} \\
\hline
$V_r$  &  $  0.3 \pm 0.1$ & $ 0.02 \pm 0.05$
       &  $  1.3 \pm 0.1$ & $ 0.64 \pm 0.01$
       &  $  0.6 \pm 0.1$ & $ 0.47 \pm 0.02$ \\

  $V$  &  $ 11.6 \pm 0.5$ & $ 0.97 \pm 0.01$
       &  $ 13.5 \pm 0.5$ & $ 0.21 \pm 0.01$
       &  $  6.7 \pm 0.5$ & $ 0.14 \pm 0.01$ \\

$b-y$  &  $  3.9 \pm 0.5$ & $ 0.02 \pm 0.02$ 
       &  $  2.5 \pm 0.5$ & $ 0.32 \pm 0.03$ 
       &  $  2.1 \pm 0.4$ & $ 0.21 \pm 0.03$ \\

$c_1$  &  $ 11.7 \pm 0.8$ & $ 0.53 \pm 0.01$ 
       &  $  7.9 \pm 0.8$ & $ 0.80 \pm 0.02$ 
       &  $  2.5 \pm 0.7$ & $ 0.63 \pm 0.05$ \\

$u-b$  &  $  5.8 \pm 0.7$ & $ 0.55 \pm 0.02$ 
       &  $  4.1 \pm 0.7$ & $ 0.85 \pm 0.03$ 
       &  $  1.7 \pm 0.6$ & $ 0.60 \pm 0.06$ \\
\hline
\end{tabular}                                                    
\end{center}
\end{minipage}
\end{table*}

\section{Radial velocity}

The radial velocity is determined from the displacement of a spectral line
from its laboratory wavelength.  If the line is symmetric, the displacement
is uniquely defined, since the centroid, median and mode coincide.  If it is
not symmetric, problems arise as to what is used in defining the
displacement.  This is always a problem in pulsating stars, since the lines
are almost always skew.  The cross-correlation technique is often used to
determine the radial velocity, but this does not get round the problem which
is now transferred to defining the position of the (possibly asymmetric)
correlation function.  In this paper we will adopt the {\em centroid} (the
first moment of the line profile) as the measure of displacement because it
is easy to calculate and because it is used in theoretical work on pulsating
stars.

In the case of $\gamma$~Dor, the most accurate measurement of radial
velocity is obtained by making use of the echelle spectra from MSSSO and
SAAO.  We used the first spectrum obtained at SAAO as the master and determined the
radial velocity by cross-correlation with this spectrum.  The periodogram of
the radial velocities (Fig. 4) shows a peak at $f = 1.365 \pm 0.002$
d$^{-1}$, very close to $f_2$.  Although the data sampling is not sufficient
to resolve the three frequencies in the periodogram, numerical tests show
that the correct amplitudes are recovered if the frequencies are given.
Since we already know the frequencies with high accuracy, Fourier
decomposition of the radial velocities allows the amplitudes and phases to
be determined: these are shown in Table 3.  The light, colour and radial
velocities for each of the three periodic components are shown in Fig.~5.
The standard deviation of one observation for the radial velocity is
found to be 0.68 km s$^{-1}$.

\begin{figure}
\epsfysize=8.3cm
\epsffile[0 79 276 466]{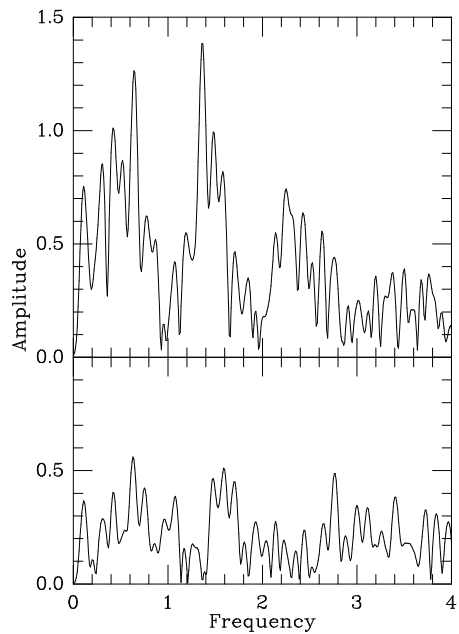}
\caption{Periodogram of the radial velocities obtained by cross-correlation
of the SAAO and MSSSO multi-order echelle data (top panel) and after removing
the frequency $f = 1.365$ d$^{-1}$ (bottom panel).  The amplitude is in
km s$^{-1}$, the frequency in cycles d$^{-1}$.}
\end{figure}

\begin{figure*}
\begin{minipage}{17.5cm}
\epsfxsize=16cm
\epsffile[74 76 452 286]{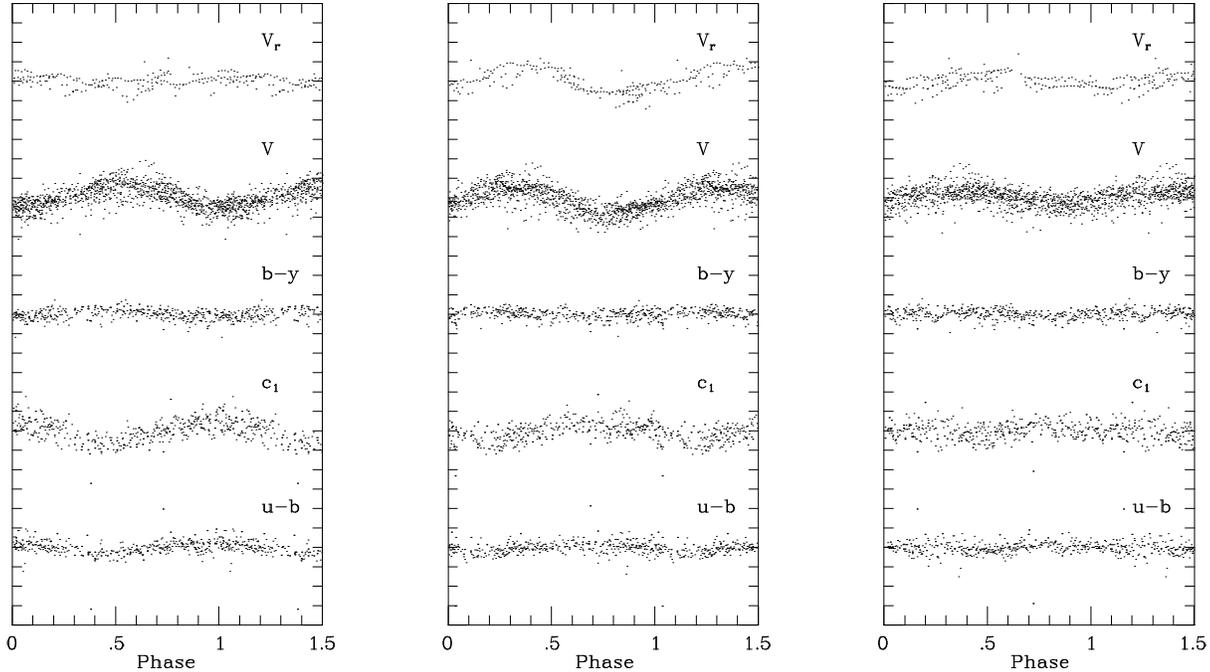}
\caption{The radial velocity, light and colour curves for the periodic
component $f_1 = 1.32098$ d$^{-1}$ (left), $f_2 = 1.36354$ d$^{-1}$ (middle)
and $f_4 = 1.47447$ d$^{-1}$ (right).  The epoch of phase zero is JD 2449000.
Tick marks are spaced at intervals of 2 km s$^{-1}$ for the radial velocity
curve and 0.02 mag for the light and colour curves.}
\end{minipage}
\end{figure*}

An important result of these observations are the ratios of radial velocity
to light amplitude for the three periodic components.  Note that the light
amplitudes for the $f_1$ and $f_2$ components are comparable, but that the
radial velocity amplitude of $f_1$ is very much smaller than that of $f_2$. 
It is not easy to understand how two starspots could produce such different
radial velocity to light amplitude ratios.  On the other hand, NRP can
account for this difference if the two components belong to different modes. 
The ratios of radial velocity to light amplitude for $f_2$ and $f_4$ are
nearly the same, indicating that ($\ell, m$) for these two components could
be the same (as indeed they are - see below).  Note also that in $f_2$,
radial velocity maximum occurs near light maximum. 
This again is not consistent with the starspot model, where you would
expect radial velocity maximum to occur when the spot is receding, i.e. a
quarter of a period after minimum light for a dark spot (or a quarter of a
period after maximum light for a bright spot).

\section{The moments}

To determine the pulsation modes for each of the three periods in
$\gamma$~Dor, we need to measure the first few moments of a particular
spectral line as a function of time.  As already mentioned, the
$\lambda$5018.450 {\AA} line of Fe~II was chosen because it is strong and
relatively unblended.  The procedure for obtaining the moments has been
automated in a program called {\tt DoSpec} (written by L Balona).

In general, {\tt DoSpec} looks for spectral absorption lines with central depth greater
than a certain threshold, fits them by Gaussians, and removes them from the
spectrum.  The threshold is lowered and the procedure repeated until a
certain level is reached.  {\tt DoSpec} then adds each fitted line back to
the spectrum, calculates the moments, and removes the line again.  In this
way moments are calculated which are relatively free of the effects of
neighbouring lines.  The output is a file listing all lines in the spectrum
and their first seven moments (moments 2 -- 6 are calculated relative to the
centroid).  The moment of order zero is just the equivalent width (EW), the
first moment is the centroid (from which the radial velocity can be
calculated), the second moment is the rms width of the line, etc. 

{\tt DoSpec} was applied to the neighbourhood of the $\lambda$5018 line
using all available spectra (ESO, Mt Stromlo, SAAO and Mt John).  The
periodogram of the radial velocity of this line from the combined data is
rather puzzling as the strongest peak is at $f = 0.734$ d$^{-1}$.  This is
different from any frequency so far identified, though it is rather close to
$f_4/2$.  However, if only the SAAO and Mt Stromlo data are used, the
periodogram is quite different, having a peak at $f = 1.359$ d$^{-1}$.  This
is clearly $f_2$, identified above from the radial velocity obtained by
cross-correlation.  The periodogram from the ESO and Mt John data alone
shows a peak at $f = 0.728$ d$^{-1}$.  These periodograms are shown in
Fig.~6.

\begin{figure}
\epsfysize=8.3cm
\epsffile[-12 79 264 505]{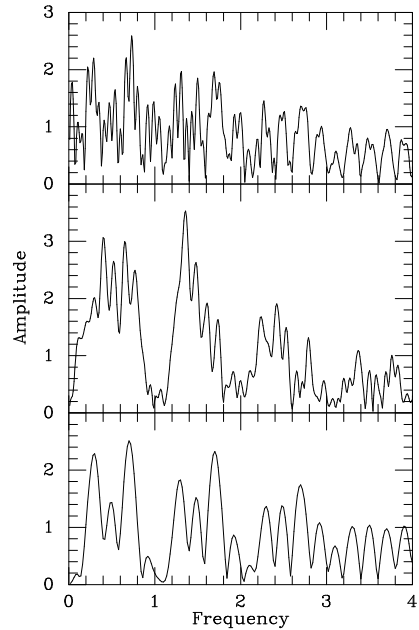}
\caption{Periodogram of the first moment of the $\lambda$5018.450 {\AA}
line of Fe~II.  Top panel - all available data; middle panel - MSSSO and SAAO
echelle data; bottom panel - ESO and Mt John data.  The amplitude is in
km s$^{-1}$, the frequency in cycles d$^{-1}$.}
\end{figure}
 
We do not understand why the ESO data do not show any power at or near $f =
1.36$ d$^{-1}$.  A phase diagram of the first moment ($m_1$) of the ESO data
phased with $f_1$, and $f_4$ (Fig.~7) shows reasonable agreement with the
SAAO \& MSSSO curves (see Fig.~8).  But for $f_2$ the curve seems
to be of the double-wave type with a minimum at about phase 0.9 and another
minimum at phase 0.4, where the SAAO \& MSSSO data show a maximum.  The
minimum at phase 0.4 arises almost entirely from data for one night
(JD~2449664).  There is no reason to suspect an instrumental cause for
this behaviour, but unfortunately we do not have photometry taken at about
the time when the double-wave form developed for $f_2$.  The star is known
to display irregular excursions in light from time to time (Balona et al.
1994a).  It is possible that such an event may be responsible for the minimum
where a maximum is expected, but this cannot be confirmed.  One should also
take into account the fact the ESO data are rather poorly sampled, several
nights having similar mean phase for $f_1$ or $f_2$.  In view of this
problem, we decided it would be better to use the SAAO \& MSSSO data only in
determining the modes.

\begin{figure}
\epsfysize=8.3cm
\epsffile[0 82 246 501]{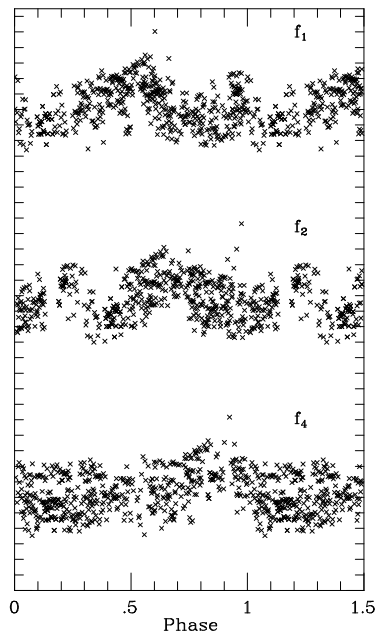}
\caption{Phase variation of the first moment of the Fe~II line
$\lambda$5018.450 {\AA} from the ESO data for the three periodic components.
The epoch of phase zero is JD 2449000.  Tick marks are spaced at intervals of
km s$^{-1}$.}
\end{figure}

\section{Mode identification}

Since the phase relationship between the light and radial velocity
variations rules out the starspot hypothesis, and since we cannot think of
any model invoking orbital motion which may account for the variations, it
is safe to assume that NRP is the most probable explanation.  However, we
need to prove that the parameters required for NRP are physically
realistic.  To do this, we need to identify the modes of oscillation and
determine these parameters.

Any arbitrary distribution can be represented by specifying its moments. 
The greater the number of moments, the more accurate is the representation. 
During pulsation, the line profile will vary and so will its moments.  For
example, in $\gamma$~Dor the moments will vary with the three frequencies,
$f_1$, $f_2$ and $f_4$.  The amplitude and phase of a particular moment
which belongs to one of the three oscillations can be calculated by
fitting a multiperiodic Fourier curve.  For a particular angle of inclination,
$i$, and a given mode, ($\ell, m$), the amplitudes and phases of the first
few moments can be fitted to the predicted ones, and the pulsation parameters
obtained by least squares.  In general, the fit will be poor unless the
chosen values of $i$, $\ell$ and $m$ are close to the true values.  In this
way, the most likely values of $i$, $\ell$ and $m$ can be estimated.

This method of mode estimation was developed by Balona (1986a, 1986b, 1987).
It uses the algorithms described in
Balona (1987) and assumes that the pulsation velocity amplitude is much
smaller than the projected rotational velocity.  This is certainly true in
$\gamma$~Dor where the radial velocity semi-amplitude is less the 2 km
s$^{-1}$, whereas $v \sin i = 50$ km s$^{-1}$.  The algorithm uses a
first-order expansion in the ratio $\Omega/\omega$ ($\Omega$ being the
frequency of rotation and $\omega$ the frequency of pulsation). The
eigenfunctions of a rotating, pulsating star are well described by pure
spherical harmonics only when this ratio is small.  This is not the case in
$\gamma$~Dor.  In Balona et al. (1994a), the stellar radius is found to be
$R = 1.32 R_\odot$.  Using the value $v \sin i = 50$ km s$^{-1}$, gives a
rotation period $P_{\rm rot} < 1.34$ d.  Therefore $\Omega/\omega > 0.55$.  For
such a large ratio, departures of the eigenfunctions from pure spherical
harmonics are bound to be significant and we do not expect a very good fit
on the basis of a first-order expansion.  Spheroidal modes with different
values of ($\ell, m$) are mixed in, and so are toroidal modes.  These
additional components will have lower amplitudes, so we can expect the method
to identify the dominant spherical harmonic mode ($\ell, m$).

Aerts \& Krisciunas (1996) apply a similar method in their analysis of
9~Aur.  However, the rotational velocity of this star is rather low and the
velocity variations  larger than in $\gamma$~Dor.  The approximation
discussed above is probably not a good one for this star, and a different
algorithm was used.  Owing to the low rotational velocity, the algorithm is
incapable of distinguishing between prograde ($m < 0$) and retrograde ($m >
0$) modes.

We use the Fe~II line $\lambda$5018.450 {\AA} to probe the pulsation mode. 
It may be possible to use more than one line to improve the signal to noise
ratio.  In fact, Aerts \& Krisciunas (1996) use the profile obtained from
the cross-correlation of the spectrum with a mask in the CORAVEL instrument
for this purpose.  We have not tested the reliability of this procedure.
First, we calculate the centroid (first moment) of the line profile, which
gives the radial velocity.  Then we calculate higher-order moments with
respect to the centroid as origin.  The moments which are actually required
are those with respect to the centre of mass of the star.  This means that
we need to determine the mean radial velocity, and then re-calculate all the
higher order moments with respect to this velocity (rather than the
centroid).  Transformation of the origin of the moments from the centroid to
the centre-of-mass velocity is easily accomplished using a polynomial
expansion.  Using only the MSSSO and SAAO data, the mean radial velocity is
$<m_1> = 22.4 \pm 0.2$ km s$^{-1}$; the standard deviation of one observation
being 2.3 km s$^{-1}$.  The standard deviation is larger than in the
cross-correlation technique because only one line is used.  A few published
radial velocity measurements are available (Abt \& Biggs 1972).  These range
from 20 to 27 km s$^{-1}$, in good agreement with the value obtained here from
the Fe~II line.

In Fig.~8, we show the variation of the equivalent width (EW)
and the first four moments of the Fe~II line $\lambda$5018.450 {\AA} phased
with $f_1$, $f_2$ and $f_4$ for the combined MSSSO and SAAO data.
The amplitudes and phases are shown in Table~4.  We have fitted the data
with all three periodicities because these periodicities are present in the
light curve.  It might be argued that the only period which is definitely
present in the radial velocities is $f_2$.  It is certainly true that $f_2$
has the highest amplitude, but this could be merely a result of the
particular nonradial mode favouring the first moment as opposed to other
moments.  The amplitude of the radial velocities (and other moments) places
a constraint on the mode identification even if it is close to zero.  It is
therefore necessary to determine the amplitudes and phases of the moments
of all three periodicities if one is to understand how the light variations
are produced.

\begin{figure*}
\begin{minipage}{17.5cm}
\epsfxsize=17.5cm
\epsffile[74 76 452 286]{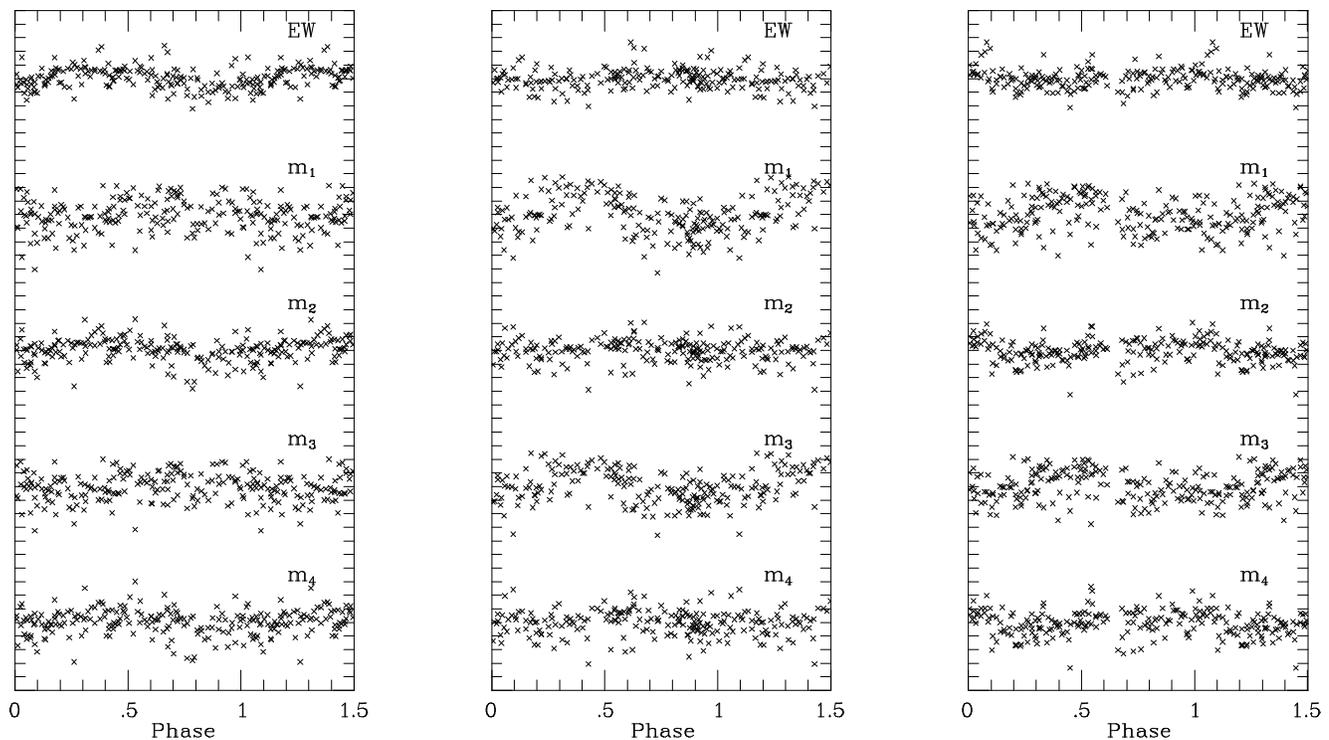}
\caption{Phase variation of the equivalent width (EW) and the first four
moments ($m_0$ -- $m_4$) of the Fe~II line $\lambda$5018.450 {\AA} phased
with $f_1$ (left), $f_2$ (middle) and $f_4$ (right).  The epoch of phase zero
is JD 2449000. Tick marks are spaced at intervals of 0.02 {\AA} for EW and
2 km s$^{-1}$ for $m_1$.  The other moments are plotted on an arbitrary scale.}
\end{minipage}
\end{figure*}

\begin{table*}
\begin{minipage}{17.5cm}
\caption{Amplitudes ($A_j$) and phases ($\phi_j$) of a three-component
Fourier fit to the EW and first four moments of the Fe~II line
$\lambda$5018.450 {\AA} for the combined MSSSO and SAAO data.  The frequencies
are $f_1 = 1.32098$, $f_2 = 1.36354$, $f_4 = 1.47447$ d$^{-1}$ and the epoch
is JD 2449000.  Semi-amplitudes are in m{\AA} for the EW, in km s$^{-1}$ for
$m_1$ and in powers of km s$^{-1}$ for the other moments ($m_2 \times 10^2,
m_3 \times 10^4, m_4 \times 10^5$).  Phases are in periods.}

\begin{center}
\begin{tabular}{crrrrrr}\hline
 & \multicolumn{1}{c}{$A_1$} &  \multicolumn{1}{c}{$\phi_1$} 
 & \multicolumn{1}{c}{$A_2$} &  \multicolumn{1}{c}{$\phi_2$} 
 & \multicolumn{1}{c}{$A_4$} &  \multicolumn{1}{c}{$\phi_4$} \\
\hline
EW     &  $  11.4 \pm 2.2$ & $  0.60 \pm 0.03$
       &  $   5.9 \pm 2.6$ & $  0.24 \pm 0.06$
       &  $   5.0 \pm 1.8$ & $  0.04 \pm 0.06$ \\

$m_1$  &  $ 1.0 \pm 0.3$ & $  0.30 \pm 0.06$
       &  $ 2.3 \pm 0.4$ & $  0.61 \pm 0.03$
       &  $ 1.4 \pm 0.3$ & $  0.53 \pm 0.03$ \\

$m_2$  &  $ 1.1 \pm 0.3$ & $  0.59 \pm 0.04$ 
       &  $ 0.4 \pm 0.3$ & $  0.38 \pm 0.12$ 
       &  $ 0.6 \pm 0.2$ & $  0.12 \pm 0.05$ \\

$m_3$  &  $ 0.7 \pm 0.3$ & $  0.27 \pm 0.07$ 
       &  $ 1.9 \pm 0.3$ & $  0.62 \pm 0.03$ 
       &  $ 1.2 \pm 0.3$ & $  0.54 \pm 0.03$ \\

$m_4$  &  $ 9.0 \pm 3.0$ & $  0.58 \pm 0.05$ 
       &  $ 4.8 \pm 3.0$ & $  0.39 \pm 0.11$ 
       &  $ 6.1 \pm 2.4$ & $  0.19 \pm 0.06$ \\
\hline
\end{tabular}                                                    
\end{center}
\end{minipage}
\end{table*}

The mean equivalent width is $<EW> = 0.369 \pm 0.001$ {\AA} (the standard
deviation for one observation is 0.015 {\AA}).  It is surprising to find that
there is significant variation of the EW when phased with $f_1$, but not with
$f_2$ or $f_4$.  For the $f_1$ component, the EW reaches a maximum at about
the time of maximum $c_1$ colour, i.e. slightly before maximum light.  On the
other hand, there is no significant variation with $f_2$ even though the light
and colour amplitude is almost the same as $f_1$.

Only a few moments show significant phase variation, which is not
surprising in view of the low radial velocity amplitude of the star.  This
means that mode identification will be indeterminate unless we use the most
significant variations and employ as many constraints as possible.  In the
case of $\gamma$~Dor we need to consider how we can constrain the pulsational
parameters.  NRP generates a variable temperature distribution on the
photosphere which distorts the line profile owing to rotation of the star. 
This effect can be quantified as a fictitious velocity field, $v_f$.  It is
fictitious because it is not generated by the variation of displacement
during pulsation.  Instead, the spatial variation of temperature across
the photosphere caused by nonradial pulsation, in combination with rotation
of the star, adds or removes flux from the line profile.  The net result is
a change in the radial velocity as measured by the centroid or line minimum. 
It plays an important role only if rotation is reasonably rapid.  Very
crudely, the amplitude of the velocity generated by this effect is
approximately given by the product of the relative luminosity variation and
the projected rotational velocity.  In $\gamma$~Dor, the semi-amplitude of
the light variation is about 0.01 mag and $v \sin i = 50$ km s$^{-1}$, so we
expect $v_f \approx 0.5$ km s$^{-1}$.  This is comparable to the observed
radial velocity semi-amplitude and cannot be neglected.
  
The other velocity fields are $v_r$ -- the pulsational velocity in the
vertical direction and $v_h$ -- the pulsational velocity in the horizontal
direction.  For each of these three velocity fields, the amplitude and phase
needs to be determined, giving six free parameters for a given $i$, $\ell$
and $m$.  We know that $v_r$ must be small compared to $v_h$ because
the three modes must all be $g$-modes.  The ratio\\
$${v_h \over v_r} \approx {4 \pi^2 f^2 R^3 \over GM},$$

\noindent
is approximately 36 for $\gamma$~Dor.  Hence we may safely ignore $v_r$ and
limit the solutions to $v_f$ and $v_h$ only, resulting in four free
parameters.  A further constraint can be imposed: $v_f < v \sin i$, since
larger values imply a relative flux variation greater than unity.

To first order, the horizontal velocity component, $v_h$, will not contribute
to the light variation since there is no geometric distortion.  All the light
variation arises from the temperature variation during pulsation ($v_f$).  On
the other hand, $v_h$ will certainly contribute to the distortion of the line
profile.  The amplitude and phase of the light curve is determined only by
$v_f$, but the amplitude and phase of the radial velocity curve depends on
$v_h$ as well as $v_f$.  This allows for a different phase relationship
between the two curves other than that predicted by the spot model.

The light curve and the radial velocity curve can therefore be used to
determine $v_f$ and $v_h$ for any $i$, $\ell$ and $m$, but at least one more
moment is required to produce a discriminant.  In principle, a solution can
be obtained using the amplitudes and phases of the light curve and the first
two moments only.  However, there is information in the third and higher
moments which should be used to constrain the solution.  It is clear, however,
that the higher order moments should be given less weight because they have
larger uncertainties.  The weights to be assigned to a particular moment can
be determined from the standard error of its amplitude and phase (easily
calculated from the Fourier fit).  The equations of condition in the algorithm
have dimensions of velocity, since each equation is divided by ($v \sin
i)^{n-1}$ for the $n$-th moment.  Hence the appropriate weight is obtained
by dividing the standard error of the moment by this quantity.  This
results in the following standard deviations being assigned to each moment:
light curve -- 1.00; first moment -- 2.09; second moment -- 2.73; third moment
-- 6.82; forth moment -- 11.84 km s$^{-1}$.

The zero points of the second and fourth moment of the line profile are
sufficient to determine the projected rotational velocity and the rms line
width ($W_i$) of the intrinsic profile.   $W_i$ includes a contribution to
the intrinsic line width owing to the finite resolution of the instruments
(the difference in resolving power of the SAAO and MSSSO spectrographs is
equivalent to about 5 km s${-1}$).  In principle, it is possible to determine
$W_i$ for the two instruments, but it is unlikely that this difference
would be significant.  The zero points of the second and fourth moments
are $m_2 = 1897 \pm 14$ and $m_4 = (10.54 \pm 0.16) \times 10^6$ in units
of km s$^{-1}$ to the respective powers.  Using a limb darkening value of
$u = 0.5$, appropriate for an early F-type star in the visible region, we
find $v \sin i = 48$ km s$^{-1}$ and $W_i = 37$ km s$^{-1}$.
The value of $v \sin i$ agrees well with that of Slettebak et al. (1975).
These values are used to deconvolve the moments to the values they would
have had in the limit of zero intrinsic width as described in Balona (1986a,
1986b, 1987).

Using these values, we calculated the discriminant, $\sigma$, for a choice of
mode ($\ell, m$) as a function of angle of inclination, $i$, using the
weighted light curve and first four moments.  In this calculation, we
adopted a radius $R = 1.32 R_\odot$, a limb darkening coefficient $u =
0.50$, and the Ledoux rotational splitting coefficient, $C = 0.1$ (a typical
value; the exact value is unimportant).  The computation included all modes
up to $\ell = 4$.  The discriminant is plotted against angle of inclination in
Fig.~9 for the three modes.

\begin{figure}
\epsfysize=8.3cm
\epsffile[0 82 273 456]{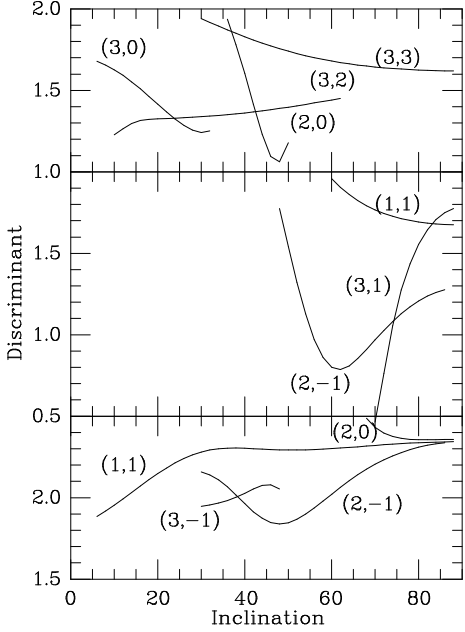}
\caption{Discriminant (arbitrary units) as a function of angle of
inclination (in degrees) for various modes ($\ell, m$).  Top panel: $f_1 =
1.32098$ d$^{-1}$; middle panel: $f_2 = 1.36354$ d$^{-1}$; bottom panel:
$f_4 = 1.47447$ d$^{-1}$.}
\end{figure}

Since the angle of inclination must be the same, $f_2$ constrains
$i > 50^\circ$.   The (3,~1) mode leads to a light amplitude which is too
low except over a very small range near $i = 77^\circ$.  Therefore, we can
identify $f_2$ as either (2,~-1) or (1,~1).  Both give realistic pulsation
parameters and good agreement with the light amplitude for an inclination
$65 < i < 80^\circ$, but the (2,~-1) mode does not match the phase of the light
curve very well.  With this limit on the angle of inclination, we find that $f_1$
can only be (3,~3).  This also gives realistic pulsation parameters and good
agreement with the light curve.  Finally, we find (1,~1) as the best estimate
for $f_4$, since (2,~0) and (2,~-1) produce light amplitudes which are twice
as high as observed.

Another way of imposing the constraint that the inclination angle for all
three modes must be the same, is to calculate the sum of the weighted
discriminant at a particular value of $i$ for the three modes for all values
of ($\ell, m$).  When the minimum of this sum is plotted against angle of
inclination, an indication of the uniqueness of the solution can be
ascertained.  This is shown in Fig.~10 where we have chosen weights of unity
for $f_1$ and $f_2$, and a weight of 0.25 for the less significant $f_4$. 
There is no solution for $i < 7^\circ$ because this would imply an
equatorial velocity larger than the breakup velocity.  The variance for
$45 < i < 63^\circ$ is too large to be plotted.  The figure shows that the
best solution is found for $i > 63^\circ$.  As mentioned above, the
identifications for this range are (3,~3) for $f_1$, (2,~-1) or (1,~1) for
$f_2$ and (1,~1) for $f_4$.

\begin{figure}
\epsfxsize=8.3cm
\epsffile[50 82 440 310]{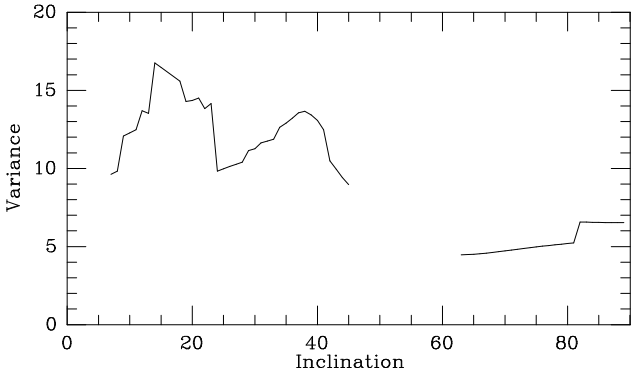}
\caption{The minimum sum of the discriminant variance over all three modes
is shown as a function of angle of inclination (in degrees).  The best
solution is found for $63 < i < 80^\circ$.}
\end{figure}

On the basis of these data, we conclude that the angle of inclination of
$\gamma$~Dor is probably $i \approx 70^\circ$, and that the best estimate of
the values of ($\ell, m$) for the three oscillations are: (3,~3) for $f_1$,
(1,~1) for $f_2$ and for $f_4$.  Table~5 gives the pulsational parameters
for this angle of inclination.

\begin{table*}
\begin{minipage}{17.5cm}
\caption{Pulsation parameters for the most probable mode identifications,
($\ell, m$), for the three oscillations in $\gamma$~Dor.  The angle of
inclination is assumed to be $i = 70^\circ$.}

\begin{center}
\begin{tabular}{crrrrrrrrrr}\hline
& ($\ell,m$) & $\sigma$ & $v_f$ & $\phi(v_f)$ & $v_h$ & $\phi(v_h)$ &
$A(V)$ & $\phi(V)$ & $A(V_r)$ & $\phi(V_r)$ \\
\hline
$f_1$ & (3, 3) & 1.64 & 15.2 & 0.13 &  8.4 & 0.09 & 0.006 & 0.13 & 1.00 &
0.31 \\
$f_2$ & (1, 1) & 1.77 &  3.2 & 0.25 & 12.8 & 0.35 & 0.016 & 0.25 & 2.35 &
0.61 \\
$f_4$ & (1, 1) & 2.32 &  1.6 & 0.26 &  8.5 & 0.26 & 0.008 & 0.26 & 1.54 &
0.51 \\
\hline
\end{tabular}                                                    
\end{center}
\end{minipage}
\end{table*}

\section{Discussion}

One of the most puzzling aspects of $\gamma$~Dor is the sudden appearance,
from time to time, of isolated absorption or emission features (Balona et
al. 1994a).  Careful inspection of all the spectra obtained during the
MUSICOS-94 campaign has failed to reveal such oddities.  The star appears to
have a normal early-F spectrum, though broadened by rotation.
There is no doubt, however, that the light
variations of the star cannot be explained entirely by three periodic
components.  There is a certain level of non-periodicity which is not yet
understood (Balona et al. 1994a, Balona, Krisciunas \& Cousins 1994b).  The
spectroscopic observations during MUSICOS-94 show evidence for this effect on
one night (JD~2449664), where the radial velocities are substantially smaller
than expected.

Our new photometric observations confirm the existence of a third oscillation,
$f_4 = 1.47447$ d$^{-1}$, previously suspected by Balona, Krisciunas \&
Cousins (1994b).  The existence of three frequencies implies substantial
differential rotation on the basis of the starspot model.  This model also
fails to account for the great differences which exist between the radial
velocity to light amplitude ratios between the two main oscillations.  But
the most serious objection to the starspot model is the fact that for $f_2$
radial velocity and light maximum coincide in phase.  One would expect a
$90^\circ$ phase difference for a starspot.  The small radial velocity
amplitude also excludes the possibility of a binary companion.  We conclude
that nonradial pulsation is the only viable explanation for the
observations.

From the light curve and first four moments of the line profile of Fe~II
$\lambda$5018.450~{\AA}, we show that $\gamma$~Dor has a high angle of
inclination, probably $i \approx 70^\circ$ and that the three modes may be
identified as $f_1 = (3,~3)$, $f_2 = f_4 = (1,~1)$ with fair confidence.
The pulsation parameters indicate quite a large value $v_f = 15.2$ km
s$^{-1}$ for $f_1$.  Now the relative flux amplitude\\
$${\Delta F \over F} = {v_f \over v \sin i}$$\\
\noindent
which gives $\Delta F/F = 0.32$ for $f_1$, implying that this mode produces
a large temperature variation $\Delta T/T \approx 0.08$.  This may account
for the significant phase dependence of the equivalent width for this mode,
whereas the other modes show little or no phase dependence.  The fact that
the amplitudes of the colour indices $c_1$ and $(u-b)$ are larger in this mode
is also consistent with the large $v_f$.

The fact that $f_2$ dominates the radial velocity variation, but that $f_1$
does not, though they have similar light amplitudes, is due to the fact that
$f_1$ has a relatively high spherical harmonic degree $\ell = 3$, whereas
$f_2$ has the lower value $\ell = 1$.  The averaging effect in the
latter case is much less, leading to a larger radial velocity amplitude.  On
the other hand, the much larger value of $v_f$ for $f_1$ compensates in the
light curve.

It is interesting that all three modes seem to be sectorial retrograde
modes.  The frequency of pulsation in the co-rotating frame, $\nu_0$, is
related to the frequency of pulsation in the observer's frame, $\nu$, by:
$\nu_0 = \nu + m \Omega$ where $\Omega$ is the frequency of rotation.
If $i = 70^\circ$ and $R = 1.32 R_\odot$, we find $\Omega = 1.04$ d$^{-1}$
($P_{\rm rot} = 0.96$ d).  Thus, in the co-rotating frame, the three oscillations
have the following frequencies: $f_{10} \approx 4.44$, $f_{20} \approx 2.40$
and $f_{40} \approx 2.51$ d$^{-1}$.  $f_1$ and $f_2$ are substantially
different and must have quite different radial orders in spite of the fact
that they are nearly equal in the inertial frame.  On the other hand, $f_2$
and $f_4$, which are of the same degree, $\ell$, probably have neighbouring
radial orders.  This information could be used to extract seismological
information, but the effect of rotation needs to be taken into account more
accurately before this can done.

The success of mode identification in this star is encouraging.  The
application of the method to other stars might well result in very
interesting results, as in the case of 9~Aur (Aerts \& Krisciunas 1996),
but other types of pulsating star may be equally interesting. 
For example, an intensive campaign on a periodic Be star (and there are many
of these which are bright) should result in a definitive test of the
NRP/starspot models (Balona 1996).  All that would be required is an
accurate determination of the radial velocity to light amplitude ratio.

\section*{Acknowledgements}
JBH thanks O. K. Petterson for assistance with MIDAS reductions.  WAL thanks
Dr S James, ADFA, for computing support.

\bsp
\end{document}